\documentclass{ws-procs9x6-cpt22}
\usepackage{siunitx}
\usepackage{accents}
\usepackage[none]{hyphenat}
\begin{document}

\newcommand{\refeq}[1]{(\ref{#1})}
\def\etal {{\it et al.}}
\def\BFSubstantial{10}
\def\BFStrong{31.6}
\def\attd{\accentset{\circ}{a}^{(d)}_{\tau\tau}} 
\def\cttd{\accentset{\circ}{c}^{(d)}_{\tau\tau}} 
\def\attdiii{\accentset{\circ}{a}^{(3)}_{\tau\tau}} 
\def\cttdiv{\accentset{\circ}{c}^{(4)}_{\tau\tau}} 
\def\supbaylimdimiiitattfowo{2 \times 10^{-26}}
\newcommand{\keVV}{\si{\kilo\electronvolt}}
\newcommand{\MeVV}{\si{\mega\electronvolt}}
\newcommand{\GeVV}{\si{\giga\electronvolt}}
\newcommand{\TeVV}{\si{\tera\electronvolt}}

\title{New physics model constraints derived from SME coefficient
  limits using IceCube astrophysical neutrino flavour data}

\author{
Carlos.~A.~Arg\"{u}elles$^{1}$,
Kareem Farrag$^{2}$, and 
Teppei Katori$^{3}$\\  
}

{
  \address{$^1$Department of Physics and Laboratory for Particle Physics and Cosmology, \\
    Harvard University, Cambridge, Massachusetts 02138, USA}
  \address{$^2$Department of Physics and Institute for Global Prominent Research, \\
    Chiba University, Chiba 263-8522, Japan}
  \address{$^3$Department of Physics, King's College London, London WC2R 2LS, UK
}

\begin{abstract}
The IceCube collaboration has set stringent limits on neutrino sector isotropic SME coefficients through the measurement of the astrophysical neutrino flavor data.
We investigate the consequences of these limits on various new physics models.
\end{abstract}

\bodymatter

\subsection*{Introduction}
Recently, the IceCube collaboration performed the first search of Lorentz violation~\cite{IceCube:2021tdn,Farrag:CPT22} using the high-energy starting event astrophysical neutrino sample~\cite{IceCube:2020wum,IceCube:2020fpi}. 
Astrophysical neutrinos mix, forming the largest neutrino interferometers~\cite{IceCube:2017qyp,Skrzypek:2023blj} as they have the highest energy ($>10~\TeVV$) and the longest uninterrupted trajectory ($\sim$1~Mpc) among all particles. Astrophysical neutrino flavor exhibit a strong sensitivity to new physics~\cite{Bustamante:2015waa} including Lorentz violation~\cite{Arguelles:2015dca}.

Although IceCube data have shown no sign of new physics, they set stringent limits to the neutrino sector isotropic SME coefficients~\cite{Kostelecky:2011gq}. 
Limits on these SME coefficients are defined where the Bayes Factor, which compares models with versus without Lorentz violation, is $>\BFSubstantial$, corresponding to a likelihood with a one in ten chance assuming equal priors.
In particular, limits of these SME coefficients exceed na\"ive Planck scale physics expectations for non-renormalizable dimension-five ($<E_P^{-1}$) and dimension-six ($<E_P^{-2}$) operators, where $E_P\sim 10^{19}$~GeV is the Planck energy.
At present, many IceCube limits depend on the assumed astrophysical neutrino production model.
Re-analysis with higher statistics data and an improved flavor identification algorithm could remove such dependencies in the near future.

Model-independent limits are obtained for the flavor-diagonal $\tau-\tau$ elements. 
For the dimension-three SME coefficient case, the effective Hamiltonian can be explicitly written in the flavor basis as 
\begin{eqnarray}
\begin{aligned}
H_{eff}\sim
\frac{1}{2E}
\cdot
\left(\begin{array}{ccc}
m^{2}_{ee}     & m^{2}_{e\mu}  & m^{2}_{\tau e} \\
m^{2*}_{e\mu}  & m^{2}_{\mu\mu} & m^{2}_{\mu\tau}\\
m^{2*}_{\tau e} & m^{2*}_{\mu\tau} & m^{2}_{\tau\tau}
\end{array}\right)
+\left(\begin{array}{ccc}
0 & 0 & 0 \\
0 & 0 & 0 \\
0 & 0 & \accentset{\circ}{a}^{(3)}_{\tau\tau}
\end{array}\right)
=V^\dagger diag(\lambda_1,\lambda_2,\lambda_3)V~.\nonumber
\end{aligned}
\end{eqnarray}
Here, the first term is the standard neutrino mass matrix, and the second matrix represents a dimension-three isotropic SME coefficient, where we set only $\tau-\tau$ element to be nonzero. 
The sum of these two matrices is then diagonalized, and the unitary matrix $V$ controls how astrophysical neutrino flavor is modified through propagation. 
Since the mass matrix has off-diagonal terms, flavor diagonal new physics operators also modify the astrophysical neutrino flavors at detection. This is a type of Quantum Zeno effect~\cite{Harris:1980zi}. 
Large nonzero $\tau-\tau$ element is equivalent to large $e-e$ and $\mu-\mu$ elements. 
This results in the electron neutrino and muon neutrino flavor states to be locked throughout their propagation. 
Since the standard astrophysical neutrino production models include only electron and muon neutrinos, this means large nonzero $\attd$ and $\cttd$ elements forbid flavor mixing of astrophysical neutrinos, contrary to the current data whose best fit requires flavor mixing. 

A wide variety of models can produce non-standard flavor mixing. For example, interactions between neutrinos and inter- or extra-galactic space can affect astrophysical neutrino flavor, for both flavour diagonal and off-diagonal new physics. Here, we investigate new physics models listed in the Snowmass21 white paper, ``Beyond the Standard Model effects on Neutrino Flavor''~\cite{Arguelles:2022tki}. 
Below, we describe how some of these new physics models can be tested with the IceCube SME limits. 
For simplicity, we focus on the real positive part of the dimension-three limit, $\Re(\attdiii)<\supbaylimdimiiitattfowo~\GeVV$.

\subsection*{New physics models}
{\it Neutrino magnetic moment} ---            
Neutrino magnetic moments, $\mu_\nu$, is order $\sim 10^{-19}\mu_B$ in the Standard Model ($\mu_B$ is Bohr magneton). Neutrino interactions with galactic magnetic field ($\sim 6\mu$G) make a potential term $\mu_\nu\cdot B$ with order $\sim 10^{-42}$~GeV, which is far below the limit the IceCube obtains from the astrophysical neutrino flavour. On the other hand, current lab limit of the neutrino magnetic moment is order  $10^{-11} \mu_B$ from the GEMMA reactor neutrino experiment~\cite{Beda:2009kx} and the Borexino solar neutrino experiment~\cite{Borexino:2017fbd}. If the order $10^{-12} \mu_B$ neutrino magnetic moment exists due to new physics processes, this can make a potential term with $\sim 10^{-35}$~GeV. This is still many order smaller than the current SME limit from the IceCube. Thus, any new electromagnetic interactions on neutrinos are unlikely to contribute astrophysical neutrino flavour mixings.

\vspace{0.3cm}
{\it Neutrino self-interaction} ---
The Universe is filled by the cosmic neutrino background. These neutrinos produce a matter potential if there is a sizable lepton asymmetry~\cite{Diaz:2015aua}. For the standard cosmological history, neutrino temperature is 1.95~K and if we assume lepton asymmetry is $\sim 0.01$, the matter potential derived from the cosmic neutrino background is $\sim 10^{-46}$ GeV, far below reachable value by astrophysical neutrino flavour measurement. However, neutrinos may have new interactions between neutrinos, called secret self-interactions by exchanging new scalar $\phi$ or vector $\phi_\mu$ mediators~\cite{Berryman:2022hds}. If neutrinos have secret interactions like  $g_{\alpha\beta}\phi\nu_\alpha\nu_\beta$ or $g_{\alpha\beta}\phi_\mu\nu_\alpha^\dagger\bar\sigma^\mu\nu_\beta$, interactions between astrophysical neutrinos and cosmic neutrino background may be enhanced. However, the matter potential or mass correction made by such interactions seem negligible~\cite{Ge:2018uhz} because the number density of cosmic neutrino is low. This means, astrophysical neutrino flavor is unlikely affected by secret interactions. 

\vspace{0.3cm}
{\it Neutrino - Dark matter interaction} --- 
Dark matter is an optical invisible substance which consists around 85\% of all matter in the universe. 
There are many candidates of dark matter if they are new particles. 
The axion-like particles (ALPs) are a class of models with new light boson particles as dark matter, but the mass is much lower than tradional weakly interacting massive particles (WIMPs), lower than 0.01~eV to as low as $10^{-24}$~eV.
In this low mass region, ALPs behaves like a field $a$~\cite{Graham:2013gfa}, and it may couple with neutrinos~\cite{Huang:2018cwo}, $g_{\alpha\beta}(\partial_\mu a)\bar{\nu}_\alpha\gamma^\mu\gamma_5\nu_\beta$. By assuming local dark matter density $\rho_{DM}=0.3$~GeV/cm$^3$, tau neutrino-ALPs coupling $g_{\tau\tau}$ is limited order $10^{-5}$~GeV$^{-1}$ from the IceCube flavor data. 
This seems a strong limit compared with current and future experimental limits. 
There are many different types of light dark matter~\cite{Farzan:2018pnk} and it may be possible to set limits on other light dark matter candidates too.

\vspace{0.3cm}
{\it Neutrino - Dark energy interaction} --- 
Dark energy is a missing energy density which consists around 70\% of all energy density in the universe. 
One candidate uses a dynamic scalar field called quintessence $\phi$ which may couple with neutrinos with coupling constant $\lambda_{\alpha\beta}$~\cite{Ando:2009ts}. Authors~\cite{Klop:2017dim} evaluate the effect of such coupling in terms of SME coefficients, $\attdiii\sim \lambda_{\tau\tau} \dot{\phi}/M_*$, where $M_*$ is the energy scale of the interaction. We are investigating the implication of our SME limit $\Re(\attdiii)<\supbaylimdimiiitattfowo~\GeVV$ on quintessence.

\vspace{0.3cm}
{\it Long range force} --- 
Neutrinos may feel a potential through unknown long-range force. Such interaction can be speculated between neutrinos and any particles, including electrons~\cite{Bustamante:2018mzu}. Assuming long-range Yukawa type coupling between tau neutrinos and electrons, the potential can be written by the mediator mass $m_{e\tau}'$ coupling constant $g_{e\tau}'$, electron number $N_e$ and distance $d$, $V_{e\tau}={g_{e\tau}'}^2N_e(4\pi d)^{-1}e^{-d/m_{e\tau}'}$. Very small coupling constant can be explored through long range force between tau neutrinos and all electrons in the visible universe. We will address the limits on long range force in further study.

\vspace{0.3cm}
{\it  Acknowledgments} --- 
Authors thank to organizers of CPT'22 meeting to hold this conference on online.

\end{document}